\documentclass[10pt,letterpaper]{article}
\usepackage[top=0.85in,left=2.75in,footskip=0.75in]{geometry}

\usepackage{amsmath}
\usepackage{amssymb}
\usepackage{array,multirow}

\usepackage{changepage}

\usepackage[utf8x]{inputenc}

\usepackage{textcomp,marvosym}

\usepackage{cite}

\usepackage{nameref,hyperref}

\usepackage[right]{lineno}

\usepackage{microtype}
\DisableLigatures[f]{encoding = *, family = * }

\usepackage[table]{xcolor}
\usepackage{soul,xcolor}
\setstcolor{red}
\usepackage{array}

\newcolumntype{+}{!{\vrule width 2pt}}

\newlength\savedwidth

\raggedright
\usepackage[aboveskip=1pt,labelfont=bf,labelsep=period,justification=raggedright,singlelinecheck=off]{caption}

\makeatletter
\renewcommand{\@biblabel}[1]{\quad#1.}
\makeatother

\usepackage{lastpage,fancyhdr,graphicx}
\usepackage{epstopdf}
\fancyheadoffset[L]{2.25in}
\fancyfootoffset[L]{2.25in}
\usepackage[percent]{overpic}
\usepackage{bm}
\DeclareMathOperator{\tr}{tr}

\begin{document}
\vspace*{0.2in}

\begin{flushleft}
{\Large
\textbf\newline{Gang Confrontation: The case of Medellin (Colombia)} 
}
\newline
\\
Juan D. Botero\textsuperscript{1},
Weisi Guo\textsuperscript{2,3},
Guillem Mosquera\textsuperscript{2,3},
Alan Wilson\textsuperscript{3},
Samuel Johnson\textsuperscript{4},
Gicela A. Aguirre-Garcia\textsuperscript{5,6},
Leonardo A. Pachon\textsuperscript{1,7*}\\
\bigskip
\textbf{1} Universidad de Antioquia, Instituto de F\'isica, Medellin, Colombia
\\
\textbf{2} University of Warwick, Coventry, United Kingdom
\\
\textbf{3} Alan Turing Institute, London, United Kingdom
\\
\textbf{4} University of Birmingham, Birmingham, United Kingdom
\\
\textbf{5} Centro Nacional de Memoria Hist\'orica, Bogot\'a, Colombia
\\
\textbf{6} 
Freie Universität Berlin, Lateinamerika-Institut, Berlin, Germany
\\
\textbf{7} guane Enterprises, Medellin, Colombia
\bigskip

%
%

* leonardo.pachon@udea.edu.co

\end{flushleft}
\section*{Abstract}
Protracted conflict is one of the largest human challenges that have persistently 
undermined economic and social progress. 
In recent years, there has been increased emphasis on using statistical and physical 
science models to better understand both the universal patterns and the underlying 
mechanics of conflict. 
Whilst macroscopic power-law fractal patterns have been shown for death-toll in 
wars and self-excitation models have been shown for roadside ambush attacks, 
very few works deal with the challenge of complex dynamics between gangs at 
the intra-city scale. 
Here, based on contributions to the historical memory of the conflict in Colombia,
Medellin's gang-confrontation-network is presented.
It is shown that socio-economic and violence indexes are moderate to highly correlated 
to the structure of the network.
Specifically, the death-toll of conflict is strongly influenced by the leading eigenvalues of 
the gangs' conflict adjacency matrix, which serves a proxy for unstable self-excitation from 
revenge attacks.
The distribution of links based on the geographic distance between gangs in confrontation leads
to the confirmation that territorial control is a main catalyst of violence and retaliation among 
gangs. 
{As a first attempt to explore the time evolution of the confrontation network,} 
the Boltzmann-Lotka-Volterra (BLV) dynamic interaction network analysis is applied 
to quantify the spatial embeddedness of the dynamic relationship between conflicting 
gangs in Medellin.
{However, the non-stationary character of the violence in Medellin during the
observation period restricts the application of the BLV model and} results suggest that 
more involved and comprehensive models are needed to described the dynamics of Medellin's 
armed conflict.

\section*{Author contributions statement}
JDB, WG and LAP designed the research. 
JDB and LAP developed the research. 
LAP and GAA-G collected, processed and generated the conflict matrix for Medellin's gangs. 
JDB, WG and LAP wrote the paper.
All authors discussed the results of the paper. 

\newpage


\section*{Introduction}
Conflict, in one guise or another, has plagued human progress since historical records began. 
Protracted conflict is a critical force in stopping societal development and meeting the Millennium 
Goals. 
The ability to understand and predict conflict can inform peacekeeping and lead to long-term prosperity \cite{Guo18}.
Since the Cold War, conflict has increasingly become subversive, trans-national, trans-genre, 
and difficult to both define and arrest. 
Armed violence is often intermixed with illegal smuggling, narcotics, financial crimes and 
criss-cross several nations. 
The Colombian conflict is an interesting case of protracted conflict, both due to its complexity in 
the changing relationship between the governments, paramilitary groups, crime syndicates, and 
guerrillas; as well as the international attention from the illegal drug trade. 
The case of Medellin is of particular interest because it is a city that has been suffering the 
presence of gangs since the 1960s with the rise of the infamous Medellin Cartel. 
Funding from drug trafficking transformed traditional gang dynamics into violent proxy territorial 
battles for the cartels. 
Simultaneously, guerrilla groups, mainly FARC, ELN and EPL, established cooperation 
pacts with the gangs, increasing their influence and ability to recruit. 
In the late 90's and early 00's the presence of paramilitary groups in the socio-political context 
of Colombia also allowed those groups to co-opt the gangs in Medellin \cite{melguizo2001evolution}. 
After the 2003-2005 demobilization agreement between the paramilitary army and the Colombian 
government, gangs come back again to the service of the narco-cartels and co-opt local 
legal economies. 

The qualitative arguments on territorial conflict at the intra-city scale have been well understood 
over the decades through extensive ground-level studies by the \textit{Instituto de Estudios Politicos-- 
UdeA}, \textit{Instituto Popular de Cultura}, 
\textit{Centro de Analisis Politico -- Universidad EAFIT} and recently the \textit{Centro Nacional de 
Memoria Histo\'orica} (CNMH). 
However, there lacks a unified quantitative model which can both describe the chain of historical 
conflict events, as well as potentially forecast future conflict dynamics. 

\subsection*{Literature Review}
Conflict prediction can generally be divided between data-driven statistical methods and causal 
reasoning \cite{Guo18}. In data-driven methods, statistical trends (e.g. cycles, power-laws\cite{Richardson41}, 
spatial-temporal processes \cite{PNAS12}) are used to guide prediction. 
Recently, the availability of high spatial resolution data allowed researchers to have significant 
impact in the field \cite{d2016crowdsourcing, van2016role} and show that the statistical patterns 
are significant \cite{Bohorquez09, Johnson13, d2015kickoff, Clauset13}, and can have self-excitation 
behaviour (i.e., Hawkes process) \cite{Fry16}. 
However, the accuracy of such models is either confined to the aggregate scale or fine-tuned to 
work in a highly specific context. Furthermore, the low-dimensional model parameters often do 
not naturally reflect the multi-dimensional causal factors. 
On the other hand, causal reasoning is used to combine domain expertise and real-time knowledge 
to predict violence \cite{Mac04}.
Groups of experts have been shown to be effective in reducing bias and surveys of experts to quantify 
risk predicts general trends well \cite{God00}.

Alternative Agent-Based Model (ABM) approaches are on the other hand able to test hypotheses 
and causal mechanisms such as policy interventions. 
Many attempts have been made to create mechanisms that explain conflict using interacting agents, 
including: clash of cultures \cite{Lim07}, distribution of political responsibility, technology transfer 
\cite{Turchin13}, foreign aid fluctuations, and deterioration of the natural environment. 
However, their complexity and data dependency means that a universal ABM is absent. 

Scalable ground census using natural language processing also works well when a curated target-specific 
learning \cite{NLP17}. 
However, such approaches do not integrate the growing data collection (e.g. ACLED, UCDP, GTD) and 
data science capabilities. 
Recently, moving beyond logistic regression, higher dimensional machine learning approaches that 
combine multiple causal factors and big data have been used to predict violence. 
Techniques such as Random Forest are able to indicate the relative importance of different factors but 
lack the mechanical insight to indicate why \cite{ViEWS19}. 
Furthermore, over-fitting and catastrophic forgetting are critical issues which will prevent the method 
from predicting unexpected new events. Indeed, even advanced deep learning techniques are likely 
to predict self-regressive behaviour (e.g., protracted war), but not new events \cite{Guo18}.

The third modelling category belongs to interaction dynamics, which attempt to model the key relationship 
dynamics between actors. 
As interactions underpins the fabric of human society across multiple population scales, methods 
such as the entropy-maximising Boltzmann-Lotka-Volterra (BLV) spatial interaction model can describe 
the projected flow of threat or influence between adjacent population groups \cite{wilson2008boltzmann}. 
Such models have been used to model ancient conflicts \cite{Baudains15} and predict the likelihood of 
new ones \cite{Guo17}.

The Boltzmann-Lotka-Volterra (BLV) model feeds from, e.g., the gravity-based principle \cite{Sto40,Zip49} 
in the social sciences that provided a formal strategy to assets the effect geographic distance 
on connectivity of spatial networks dynamics and human behaviour \cite{Sto40,Zip49,Isa54,Zip46,BSB11,
Bra12}.
The idea that the likelihood of a relationship (e.g., social or economic) is inversely 
proportional to the physical distance between two entities (Refs.~\cite{Sto40,Zip49}). 
In economic geography, the gravity model was used to explain migration flows between countries, 
regions, or cities \cite{Zip46}, and showed that movement of people between cities is proportional 
to the product of their population size and inversely proportional to the square of the 
distance between them. 
In the context of international economics, the gravity model of trade predicts trade- flow volumes 
and capital flows between two units to be directly proportional to the economic sizes of the units 
(using GDP data) and inversely proportional to the distance between them \cite{Isa54}. 

In the context of corporate competition, spatial network analysis was utilized to show \cite{BSB11} that 
the spatial locations of firms are positively correlated with the population density, and that firm 
competition networks are governed by cumulative advantage rules and geographic distance (which is 
equivalent to the BLV).
In the contexts of civil unrest and riots \cite{Bra12,BB&18}, it has been shown, both theoretically 
and empirically that social unrest dynamics is based on the hypothesis that widespread unrest 
arises from internal processes of positive feedback and cascading effects in the form of contagion 
and social diffusion over spatially interdependent regions connected through social and mass 
communication networks.
So that social instability can be considered as a spatial epidemics phenomenon, similar to other 
spatially extended dynamical systems in the physical and biological sciences, such as earthquakes, 
forest fires, and epidemics. 
This perspective was confirmed by modelling the 2005 French riots using spatial epidemiological models
\cite{BB&18}.
Here we extend these ideas to the case of urban paramilitary groups in their hegemonization 
process in Medellin, Colombia.

\subsection*{Contribution}
The gang confrontation network, over twenty-years of intense conflict, of Medellin is presented. 
It is shown that the violence escalation is highly correlated with socio-economical indexes like 
the Gini Coefficient, Human Right Violations, Homicide Rate and Unemployment Rate.
A high correlation between the structure of the gang confrontation network and the escalation 
of conflict in Medellin is presented. 
The collected data was analysed under the light of network theory and models 
from complex systems were employed to simulate the structure of the network. 
Specifically, application of the Boltzmann-Lotka-Vollterra model confirmed that the conflict 
network of gangs in Medellin is spatial and therefore in strongly driven by territory control.
However, results also suggest that more involved and comprehensive models are needed 
to described the dynamics of Medellin's armed conflict.

This document is organized into four sections. 
The Introduction reviews the context of the gangs in Medellin and the mathematical models implemented 
previously in a similar context.
Materials and Methods Section discusses the data sources and presents the dynamic analysis 
of networks and the BLV formalism. 
Results and Discussion Section covers the main results obtained from the gang's conflict network, 
the relation between socio-economic and network properties with the escalation of violence and 
the simulations obtained after the implementation of BLV methodology. 
Finally, in Conclusions Section, the results are summarized.

\section*{Materials and Methods}
\label{sec:methods}
\subsection*{Data Sources}
After the demobilization of paramilitary groups, Colombia government created the \textit{Centro 
Nacional de Memoria Hist\'orica}--CNMH (National Center for Historical Memory) to
collect and process the contributions from demobilized people to the historical memory of the conflict.
The CNMH reconstructed the memory of Medellin conflict in the Law 1424 
Historical Memory Report on the Bloque Metro, Bloque Cacique Nutibara and Bloque Heroes
de Granda that are paramilitary structures that operated in Medellin.
In the framework of that report, information on the ego of gangs and their relationships were identified 
and processed for six well defined periods of time: (i) previous to the incursion of the Bloque
Metro, ca. 1995-2000; (ii) during the presence of the Bloque Metro, ca. 2000-2002; (iii) during
the war of the Bloque Metro and the Bloque Cacique Nutibara that annihilated the Bloque Metro,
ca. 2002; (iv) during the presence of the Bloque Cacique Nutibara that demobilized
in 2003, ca. 2003 (v) during the presence of the Bloque Heroes de Granada, ca. 2003-2008 
that demobilized in 2005 and (vi) during the Demobilization, Disarmament and Reintegration (DDR)
to civil life period, ca. 2008-2014.

Specifically, the ego of the gang comprises information on the participation in paramilitary groups, 
the illegal economies they controlled, the area of influence and an approximated number of members.
Due to the confidential character of the information, gangs were labeled with a unique code with no 
more information than a label for the administrative zone of the city where they operated and 
a random number, e.g., CE026 denotes a gang in the Center-Eastern zone (CE) and 026 is a random 
number associated to that gang in that zone.
As expected, gangs come from and operate in the most conflicted zones of Medellin's, namely, 
Center-Eastern (CE), North-Eastern (NE), Center-Western (CW) and North-Western (NW)
zones.
The information on the type of (i) confrontation, (ii) collaboration among gangs, (iii) godfathership, 
(iv) subservience and (v) types of confrontation and collaboration between gangs and State Agencies 
were registered for completeness. 
The data set was complemented with information from local media. 
For the present analysis, in the framework of the Cooperation Agreement between CNMH and 
Universidad de Antioquia (UdeA), only the dataset associated to confrontations is utilized. 

Confrontations are characterized by successive and systematic acts of direct violence, 
or escalated during periods of latency and expressed through one or successive acts of direct violence 
of different types between gangs over a period. 
Violence actions include shootings, harassment, homicides, forced 
displacement, threats; among a long list of violent actions that involve gang members and/or 
civil population of the territory under their illegal armed control.
Therefore, confrontations refer to the existence of enmity relationships or armed antagonism between 
gangs and not to a specific violent act.
Medellin's armed-conflict-nature suggested the formulation of three confrontation main categories: 
(1) direct conflict between two gangs acting by themselves, (2) conflict between a gang pertaining 
to a paramilitary or guerrilla group and a gang acting by itself and (3) conflict between gangs pertaining 
to different paramilitary or guerrilla groups.
The reasons for igniting a particular confrontation were registered as: (A) Interpersonal, (B) Territory 
defense and control, (C) Control of micro economies, (D) Control of macro economies, (E) Loyalty 
to a macro structure, (F) Counterinsurgency, (G) Self-defense and/or (H) Drug trafficking.

For instance, the confrontation between CE026 and CN032 may be characterized, e.g., as 1A, 1B.
From the information collected, five directed and weighted networks were constructed with the nodes 
being the gangs and State Agencies.
For the confrontation network, edges start in the node that ignites the confrontation and end at the 
nodes upon which the action rests. 
The weight of the edges corresponds to the number of confrontation codes needed to characterized the 
type of confrontation between two gangs.
In the example above, if CE026 ignited the confrontation against CN032 and the confrontation was 
characterized as 1A, 1B (two confrontation codes); then, the edge starts in CE026, ends in CN032 
and has weight 2.
The networks for each period were constructed, as previously done, e.g., in Ref.~\cite{BSB11}, using 
the ``snowball sampling" \cite{LKJ06}.

The CNMH-UdeA collaboration identified 671 gangs in the city across the six periods. 
During the observation periods, some gangs were annihilated and new gangs were created so that
nodes may change from period to period. 
Of the total of gangs identified in the city, information was found only for 317 of them. 
The reasons for the lack of information, in particular, for the last two periods was the apprehensiveness
of the demobilized people to contribute with information on illegal activities once they were officially 
reintegrated into the legal-civil society.
Moreover, due to the intricate nature of the conflict, it was not always possible to infer the direction 
of the edges; in these cases, undirected links were utilized.
The data was independently collected and processed by seven social scientists from ca. 70\% 
of the officially demobilized people that operated in Medellin and under the supervision of the 
lead team of CNMH and UdeA.
The data was then shared and confronted by all the members of the collaboration to agree, e.g., 
on the number of confrontation codes needed to characterize each confrontation and to deliver 
a first unified version of the network.
A comprehensive quantitative analysis of the networks and gangs information is in progress.

\textbf{Data availability}: The information about the number of violence acts are taken from the 
\textit{Observatorio Nacional de Memoria y Conflicto} of CNMH and accounts for information 
on infringements of International Humanitarian Law, namely, war actions, selective 
assassination, terrorist attack, damage to civilian property, enforced disappearance, massacre, 
recruitment, kidnapping, sexual violence. 
The Gini coefficient, Unemployment and Homicide Rates were compiled by the authors from open 
data provided by the \textit{Departamento Administrativo Nacional de Estadistica of Colombia} (DANE).
The datasets generated and/or analysed during the current study are attached as supporting 
information (see, S1 Table, S2 Script and S3 Script).

\subsection*{Spectral Analysis: A Dynamic Analysis Motivation}
The presence of paramilitary structures in Medellin can be conceived as a dynamical network 
with the main distinctive stages depicted by the six periods described above. 
However, the annihilation and emergence of dominant structures of very different 
character, namely, anti-insurgent (Bloque Metro), narco-paramilitary (Bloque Cacique 
Nutibara) and political-paramilitary (Bloque Heroes de Granada) suggest a separate
analysis to uncover similarities and differences in their \textit{modus operandi}.
Thus, instead of directly addressing the dynamics of the network across every stage, 
the global properties of each network are characterised below by means of a
spectral analysis of the confrontation adjacency matrix.

The eigenvalues of the adjacency matrix can be clearly related to the dynamics of the
conflict. 
In doing so, define a state vector $\bm{P}$ with components $\{ p_i\}$. 
The linearised dynamics satisfy
$
 \label{eq:dynamic}
 \dot{p}_i(t)=-p_i(t)+\sum_{j} a_{ij}p_j(t),
$
where $\mathsf{A} = a_{ij}$ is the adjacency matrix. 
$a_{ij}$ contains the information about existence of confrontation 
among the $i^{\mathrm{th}}$ and $j^{\mathrm{th}}$ nodes (gangs) whereas 
$p_k$ quantifies the intensity of the violence exerted or suffered by 
the $k^{\mathrm{th}}$ node.  
In matrix notation, $\bm{\dot{P}}(t)=-\bm{P}(t)+\bm{P}(t)\mathsf{A}^\top$.
To decouple this set of equations, note that it can be written as 
\cite{jirsa2004will}
\begin{equation}
 \label{eq:dynamicDecoupled}
 \dot{x}(t)=-x(t)+\lambda x(t), 
\end{equation} 
where $\bm{x}=\bm{P}\mathsf{e}$ and $\lambda$ and $\mathsf{e}$ being the 
eigenvalues and eigenvectors matrix of $\mathsf{A}^\top$, respectively. 
Since networks considered here are undirected, i.e., $\mathsf{A}$ is 
symmetric and real, then its eigenvalues are real.
Therefore, the system described by the differential equation 
(\ref{eq:dynamicDecoupled}) reaches a stable regime only for $\lambda<1$; 
for other values of $\lambda$, solutions diverge in the long-time 
regime.
Moreover, the Perron-Frobenius theorem \cite{berman1994nonnegative} guarantees that $\mathsf{A}$ 
will have a unique positive leading eigenvalue $\lambda_{\mathrm L}$ and the dynamics of the system 
will be mainly governed by this dominant eigenvalue of $\mathsf{A}$. 
The second largest eigenvalue $\lambda_{{\mathrm L}_2}$ can be interpreted as a second order 
correction in the stability analysis rate of convergence to equilibrium distribution.

The war rules of each period are assumed to be encoded in the adjacency matrix. 
Thus, the assumption here is that during each period the characteristics of the 
linearised dynamics are governed by the adjacency matrix.
This assumption, not verified here due to the lack of information for every
period, is then utilized to compare the global properties between all stages of 
Medellin's  conflict.

\subsection*{Boltzmann-Lotka-Volterra Models for Conflict Networks}

{As a first attempt to find an analytical model for describing the
confrontation network in Medellin, consider the Boltzmann-Lotka-Volterra (BLV)}
that have been widely used to model spatial networks 
\cite{baudains2016conflict,wilson2016global, wilson2016approaches,wilson2008boltzmann}. 
This approach applies when the external dynamics that may trigger conflicts (i.e., climate change 
and drought) are quasi-static over the time period of a few decades \cite{Hsiang2013}.  
Being BLV method a benchmark model in the field,  it is relevant to see how it 
performs here although since the data may be unlikely to be stationary. 
However, this provides a baseline for opening new lines of research that can improve on our initial 
findings.

The main goal of the BLV formalism is to merge two well-known models in science: the maximisation of
entropy proposed by Boltzmann and a competition model also known as predator-prey model proposed 
by Lotka and Volterra. 
The target of this formalism is to predict the values of the ties among the nodes 
that constitute the network, i.e., to generate the adjacency matrix. 
The values of $a_{ij}$ predicted by the theory will be bounded between $0$ and $1$, thus generating a 
weighted network that should maximise the entropy functional  $S=-\sum_{ij}a_{ij} \log a_{ij}$ with 
$\sum_{ij} a_{ij} d_{ij}=C$ and $\sum_{ij}(a_{ij} \log p_i+a_{ij} \log p_j)=B$.
Here, $d_{ij}$ is the distance between the nodes $i$ and $j$ and $p_i$ is the benefit associated to 
the $i^{\mathrm{th}}$-node, $C$ and $B$ are constants that can be understood as the total spatial 
cost and the total benefit, respectively.
Note that the BLV formalism is, therefore, a methodology to find the optimal solution of a cost-benefit 
problem.

By solving the constrained optimisation problem established above, the weight of the links between the 
nodes, as a generalisation of the Boltzmann probability distribution, can be obtained  from 
\begin{equation}
\label{eq:genAij}
a_{ij} = \frac{(p_i p_j)^{\alpha} e^{-\beta d_{ij}}}{\sum_{ij} (p_i p_j)^{\alpha} e^{-\beta d_{ij}} },
\end{equation} 
where $\alpha$ and $\beta$ are the Lagrange multipliers required for solving the constrained 
optimisation problem. 
A similar approach, but in the context of corporate competition, was performed in  Ref.~\cite{BSB11} 
by Braha \emph{et al.} 
They proposed a model combining preferential attachment and geographic distance effects, 
where the probability of competition between firms will be proportional to $(k_i k_j)^\gamma  d_{ij}^\delta$.  
The key idea is that the physical distance between nodes strongly determines the benefits and costs  
of transport and communication\cite{BSB11}, which then have significant importance in the context of 
gang war and company competition.

With no loss of generality, $\alpha$ is set to 1 in Eq.~(\ref{eq:genAij}) and the degree-preference 
model is assumed, i.e., $p_i \rightarrow k_i$, where $k_i$ is the degree of the $i^{\mathrm{th}}$-node 
of the network obtained from the data. 
Hence, the links can be written as
\begin{equation}
\label{eq:Aij}
 a_{ij} = \frac{(k_i k_j) e^{-\beta d_{ij}}}{\sum_{ij} (k_i k_j) e^{-\beta d_{ij}} }.
\end{equation}
The goal is to obtain the optimal value of $\beta$ in Eq.~(\ref{eq:Aij}) that best 
fits the confrontation network reconstructed from the collected data.
The predicted network is generated as follows: (i) the weighted edges are generated according 
to Eq.~(\ref{eq:Aij}); (ii) only the $N_\mathrm{ed}$ biggest values are selected-- $N_\mathrm{ed}$ 
is the number of edges in the real network--; and (iii) the weights are set to $1$ and the rest of them 
equal to $0$.
This was done to assure that both, the real and generated adjacency matrices, have the same 
number of links and are binary matrices.

The accuracy of the adjacency matrix obtained from the model is measured by the distance, in 
the space of matrix, to the the adjacency matrix reconstructed from the collected data. 
The distance can be calculated, e.g., by means of the Frobenius distance 
$ D_\mathrm{F}=\sqrt{\tr(A^\mathrm{r}-A^\mathrm{m})(A^\mathrm{r}-A^\mathrm{m})^\mathrm{T}}$, 
multiplying distance $D_\mathrm{M}=1- N^{-1}\sum_{ij} A^\mathrm{r}_{ij}A^\mathrm{m}_{ij}$ or the 
subtraction distance $D_\mathrm{S}=N^{-1} \sum_{ij} |A^\mathrm{r}_{ij}-A^\mathrm{m}_{ij}|$.
$A^\mathrm{r}$ is the adjacency matrix reconstructed from demobilized people contributions 
and $A^\mathrm{m}$ the adjacency matrix obtained from the model.
Finding the value of $\beta$ in Eq.~(\ref{eq:Aij}) that minimises the matrix distance 
allows for reconstructing the conflict interaction network using the BLV methodology.

\section*{Results and Discussion}
\subsection*{General Description of the Confrontation Network}
\begin{figure}[!h]
  \begin{center}
    \begin{tabular}{c}
    \includegraphics[width=0.8\textwidth]{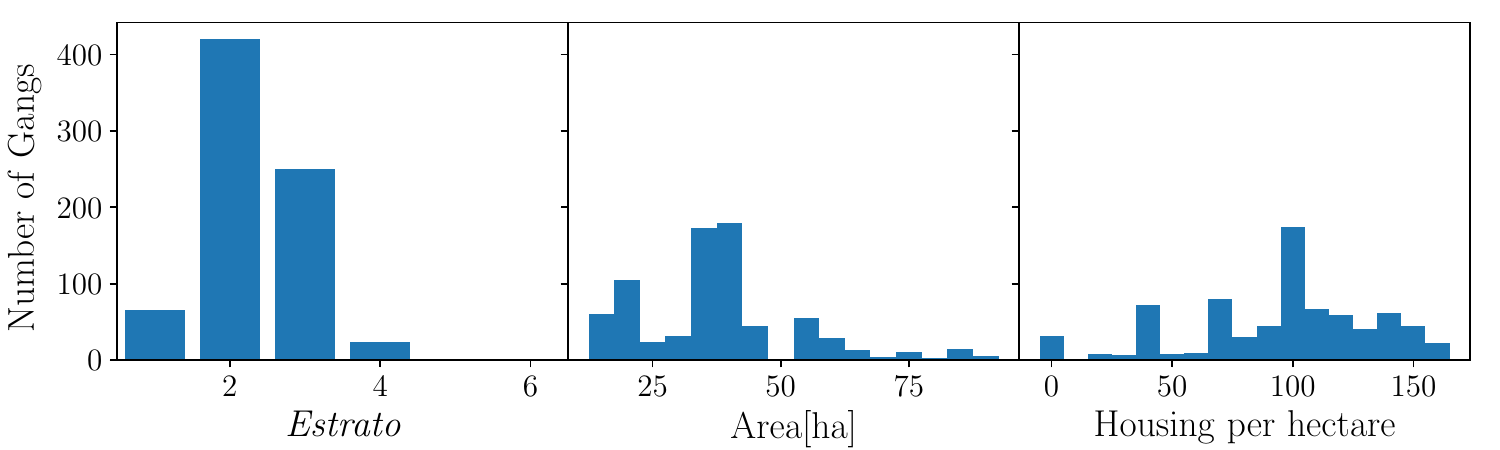}
   \end{tabular}
   \end{center}
 \caption{Left panel: Number of gangs per \textit{Estrato}. 
 Central panel: Comparison between the number of gangs and the size area per neighboorhood. 
 Right panel: Comparison between the number of gangs and the density of housing per neighboorhood}
  \label{fig:gangsVS}
\end{figure}

The distribution of gangs in the city is sociologically understood in terms of the late colonisation 
of the city, mainly, by victims of forced displacement from the countryside of Antioquia department.
Moreover, In Colombia, cities are divided into \textit{Estratos} from one to six. Citizens who live in the lowest 
\textit{Estratos}-- one, two and three-- have reduced-fares for public services (water, gas, electricity). 
Citizens who live in the highest \textit{Estratos}-- five and six-- covers the reduction for the lowest 
\textit{Estratos}.
\textit{Estrato} four corresponds to middle class, citizens who pay what they consume.
To quantitatively justify the distribution of gangs in the city, it has been commonly assumed that
the distribution and density of gangs obey geo-economics criteria.
To test the hypothesis above, Fig~\ref{fig:gangsVS} presents the number of gangs as a function of (i) 
the \textit{Estrato} of the neighboorhood, (ii) the area of the neighboorhood and (iii) the housing density per hectare.
The main presence of gangs in the lowest \textit{Estratos} can be socio-economically understood.
However, the dependence on the area of the neighboorhood and the housing density per hectare
suggests that there is no trivial explanation for the presence of gangs in different zones of the 
city.
For the case of corporate competition, in Ref.~\cite{BSB11}, it was demonstrated that the spatial locations 
of firms are positively correlated with population density. 
Although this result is intuitively clear, it cannot be generalized straightforwardly to the case of
gangs in Medellin because of its topography: a small valley that accommodates people homogeneously 
so that different the zones of the cities are fairly equally populated [Colombia official 
census 2005 by \textit{Departamento Administrativo Nacional de Estadistica of Colombia} (DANE)'s].
Therefore, to fully understand the gang phenomenon in Medellin from a geographic perspective, a
further socio-demography analysis is needed and will be conducted elsewhere.

\subsubsection*{Tolopogy of Confrontation Network}
Since it was not possible to clearly identify the area of operation of all gangs, below, two 
datasets are considered: (i) The weighted full network that contains information of 317 gangs.
For this case, the 1995-2000 network comprises 186 nodes and 277 links, in shorthand notation 
(186:277), the 2000-2002 network comprises (127:148) whereas the 2002 network (91:122) and  
the 2003,  2003-2008  and 2008-2014 networks have (83:131), (40:48) and (80:119),
respectively.
(ii) The unweighted geo-referenced network comprises the gangs for which it was possible 
to collect georeferencing information; additionally, this network considers no difference between 
having one or several reasons to get into conflict.
The topology analysis is performed only for the full network whereas the subsequent 
analysis is comparatively presented, when possible, for both networks.
Information needed to reproduce the results presented here is provided in the Supplementary 
Information.

\begin{figure}[!h]
  \begin{center}
    \includegraphics[width=0.8\textwidth]{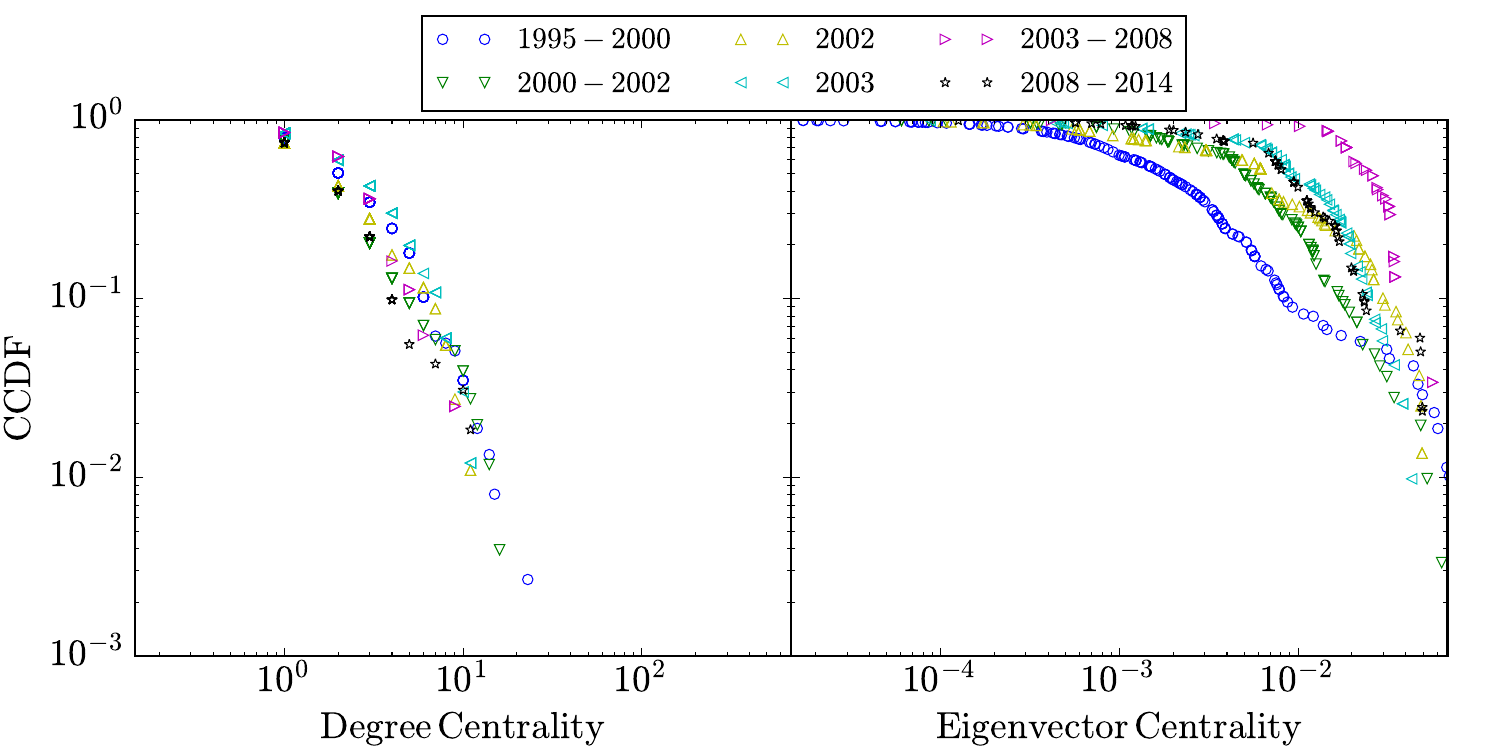}
   \end{center}
 \caption{The log-log plots of the Complementary Cumulative Distribution Function (CCDF)
  of nodes centralities.
The left hand side panel corresponds to the degree centrality measure whereas the right
hand side panel corresponds to the eigenvector centrality measure.
 }
  \label{fig:CPFAdjMat}
\end{figure}
\begin{table}[!h]
\begin{tabular}{c||ccccc}
\hline\hline
 \multicolumn{6}{c}{BIC for Degree Centrality Distributions Models} 
\\ \hline \hline
\multirow{2}{*}{Network}  & Stretched Exp. & Exp. & Power & Power Cutoff & LogNormal
\\
  & $k_{\mathrm{c}} x^{\beta_{\mathrm{c}}-1} \mathrm{e}^{-\lambda_{\mathrm{c}} x^{\beta_{\mathrm{c}}}}$ 
  & $k_{\mathrm{c}} \mathrm{e}^{-\lambda_{\mathrm{c}} x}$ 
  & $k_{\mathrm{c}} x^{-\gamma_{\mathrm{c}}}$ 
  & $k_{\mathrm{c}} x^{-\gamma_{\mathrm{c}}} \mathrm{e}^{-\lambda_{\mathrm{c}} x}$ 
  & $k_{\mathrm{c}} x^{-1} \mathrm{e}^{-\frac{(\ln x -\mu)^2}{2\sigma^2}}$ 
\\ \hline
1995-2000 & -576.259  & -\textbf{581.441} &  4729.93 & -579.259 & 341.53  \\
2000-2002 & -276.601  & -\textbf{281.462} & 210.807 & -277.87 & 231.971 \\
2002      & -293.043  & -\textbf{297.582} & 153.215 &  -293.789 & 169.178\\
2003      & -238.784 & -\textbf{243.113} & 138.868  & -239.101 & 157.503\\
2003-2008 & -77.3534 & -\textbf{81.0855} & 72.8323 & -77.4182 & 77.4182  \\ 
2008-2014 & -180.69 & -\textbf{185.081} & 136.377 & -181.201 & 151.169 \\ 
\hline \hline
 \multicolumn{6}{c}{BIC for Eigenvector Centrality Distributions Models} 
\\ \hline \hline
Network  & Stretched Exp. & Exp. & Power & Power Cutoff & LogNormal
\\ \hline
1995-2000 & -768.546  & -763.403 & 4776.1 & \textbf{-770.574} & 364.752  \\
2000-2002 & -602.729 & -\textbf{606.312} & 3288.79 & -602.956 & 233.429 \\
2002      & -290.716   & -257.489 & 98.2406 & -\textbf{293.994} & 176.977 \\
2003      & -298.711 & -\textbf{301.261} & 82.0277 & -299.546 & 159.457 \\
2003-2008 & -39.6239 & -\textbf{40.7717} & 39.4389 &  -40.4037 & 83.023\\ 
2008-2014 & -298.088 & -292.277 & 71.6443 & -\textbf{299.771}  & 149.643  \\ 
\hline \hline
\end{tabular}
\caption{Bayesian Information Criterion (BIC) for various fitting models of the 
Centrality CCFCs. 
$k_\mathrm{c}$ is calculate from 
$\int_{x_\mathrm{min}}^\infty \mathrm{d} x k_\mathrm{c} f(x)=1$ with $x_{\mathrm{min}}$ 
the the lower bound of the range of possible values that the random variable can attain.
BIC is defined as $k \ln n - 2 \ln L$ with $k$ being the number of parameters
estimated by the model, $n$ number of data points  in $x$, $x$ the observed data
and $L$ the maximised value of the likelihood function of the model.
The Model with the lowest BIC is preferred.
}
\label{tab:fitCPF}
\end{table}
For the six periods introduced above and under the assumption that the full network is undirected, 
Fig~\ref{fig:CPFAdjMat} depicts the cumulative probability function for two different centrality 
measures, namely, degree and eigenvector centralities.
The Complementary Cumulative Distribution Function (CCDF) for the six periods was fitted to 
five known distribution functions, namely,  stretched exponential, exponential, power, power 
with cutoff and LogNormal functions. 
Results are displayed in Table~\ref{tab:fitCPF}.
Based on the Bayesian Information Criterion (BIC), it is concluded that the CCDF can be accurately 
described by an exponential model $\mathrm{e}^{-\lambda_{\mathrm{c}} x}$.
Table~\ref{tab:prmtrsfitCPF} presents the relevant parameter information for the fit.

\begin{figure}[!h]
  \begin{center}
    \includegraphics[width=0.8\textwidth]{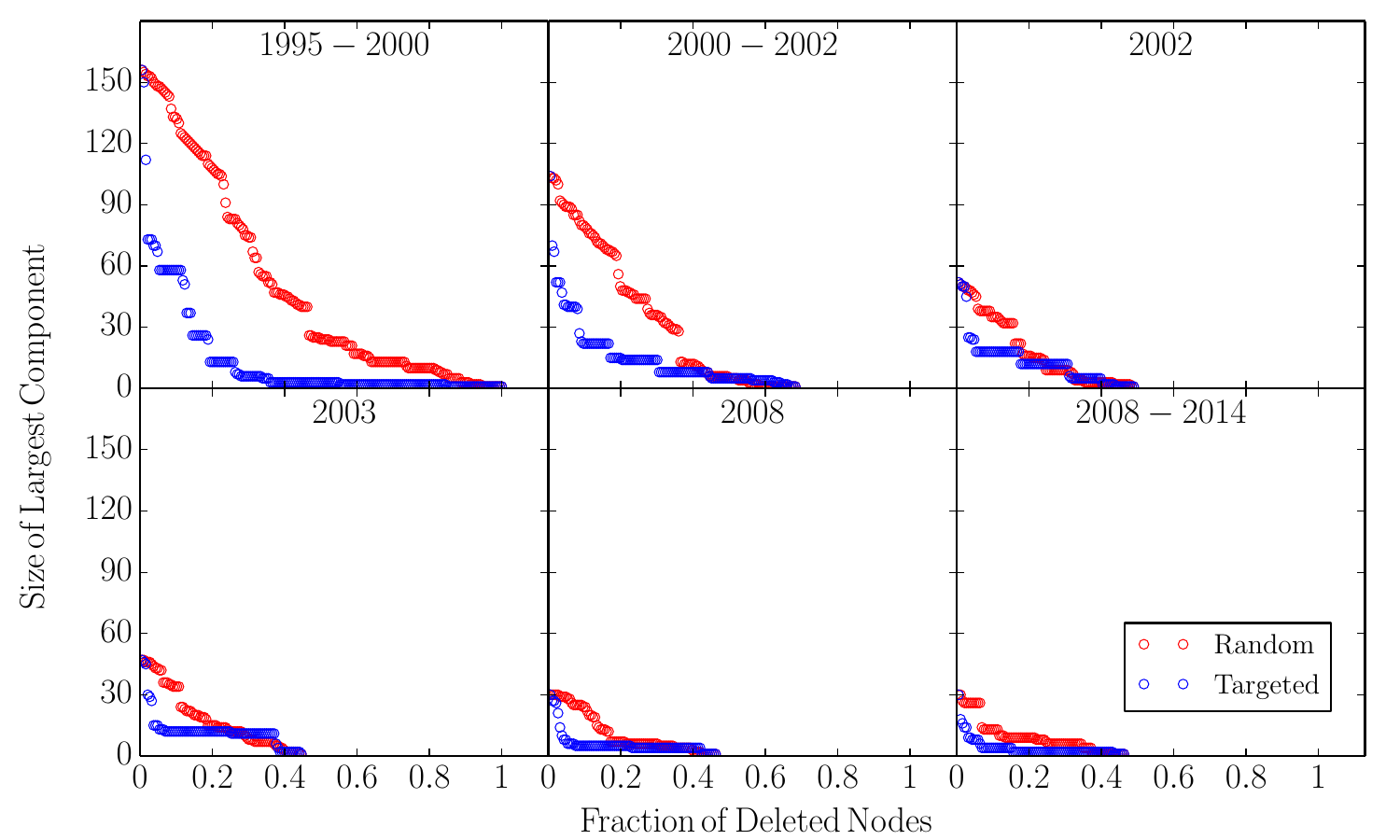}
   \end{center}
\caption{Robustness of the confrontation network, as demonstrated by the deletion of nodes. 
 Red open circles (upper) shows the size of the largest strong component surviving as nodes 
 are randomly deleted; Blue open circles (lower) shows the size of the largest component as 
 nodes are deleted in order of decreasing degree centrality.}
  \label{fig:resilience_networks}
\end{figure}

\begin{table}[!h]
\begin{tabular}{c||cc|cc}
\hline \hline
 \multirow{2}{*}{Network} & \multicolumn{2}{c|}{Degree Centrality} 
 & \multicolumn{2}{c}{Eigenvector Centrality}  
\\ \cline{2-5}
 & \multicolumn{2}{c|}{$\lambda_{\mathrm{c}}$} 
 & \multicolumn{2}{c}{$\lambda_{\mathrm{c}}$} 
\\ \cline{2-5}
 & EstValue & StError & EstValue & StError
\\ \hline
1995-2000 & 0.315704 & 0.00399511 & 365.557 & 3.38807
\\
2000-2002& 0.39655 & 0.00914728 & 136.155 & 0.967632 
\\
2002     & 0.365193 & 0.00616567 & 105.908 & 2.65042 
\\
2003     & 0.2711 & 0.00548173 & 73.0748 & 1.06943
\\
2003-2008& 0.300517 & 0.0127068 & 29.7294 & 1.95771 
\\
2008-2014& 0.401977 & 0.0108252 & 82.2071 & 1.18253
\\
\hline \hline
\end{tabular}
\caption{Fitting parameters of the CCDF in Fig~\ref{fig:CPFAdjMat} to the exponential function
$\mathrm{e}^{-\lambda_{\mathrm{c}} x}$.
}
\label{tab:prmtrsfitCPF}
\end{table}
As in Ref.~\cite{BSB11}, the robustness of the confrontation network is analysed below in terms of 
the size of the largest of the network after (i) removing randomly nodes of the network or (ii) removing
the nodes in order of decreasing degree centrality.
A network is consider fragile if after excising a single node it falls apart into many small pieces. 
The global behaviour observed in Fig~\ref{fig:resilience_networks} is that networks are resilient 
against removing random nodes, but weak against the targeted deletion of high--degree nodes.
Thus, the confrontation network of gangs in Medellin is fragile against removing highly connected
nodes.

\subsection*{Socio-economic Context of Medellin's Gang Confrontation Network}
\label{sec:results}
In terms of violence activity, there are two recent significant peaks in Medellin: one in 2002 and the 
other in 2010, which split the history into three significant periods. 
The first period (1995-2002) is related to the incursion of the paramilitary army in the city
that lead to an upsurge in violence among the {left-
and right-wing gangs}. 
This period ended with the peace process of the Colombian government with the paramilitary 
groups, a process that concluded with the demobilisation of 31.671 people in the whole country and 
corresponded to a significant reduction of Human Rights Violations until 2008. 
Between 2009-2010 other illegal armed groups such as El Clan del Golfo and La Oficina vied 
to take control over the illicit business in the city which led to the second wave of recrudescence 
of violence. 
\begin{figure}[!h]
  \begin{center}
    \begin{tabular}{c}
    \includegraphics[width=0.9\textwidth]{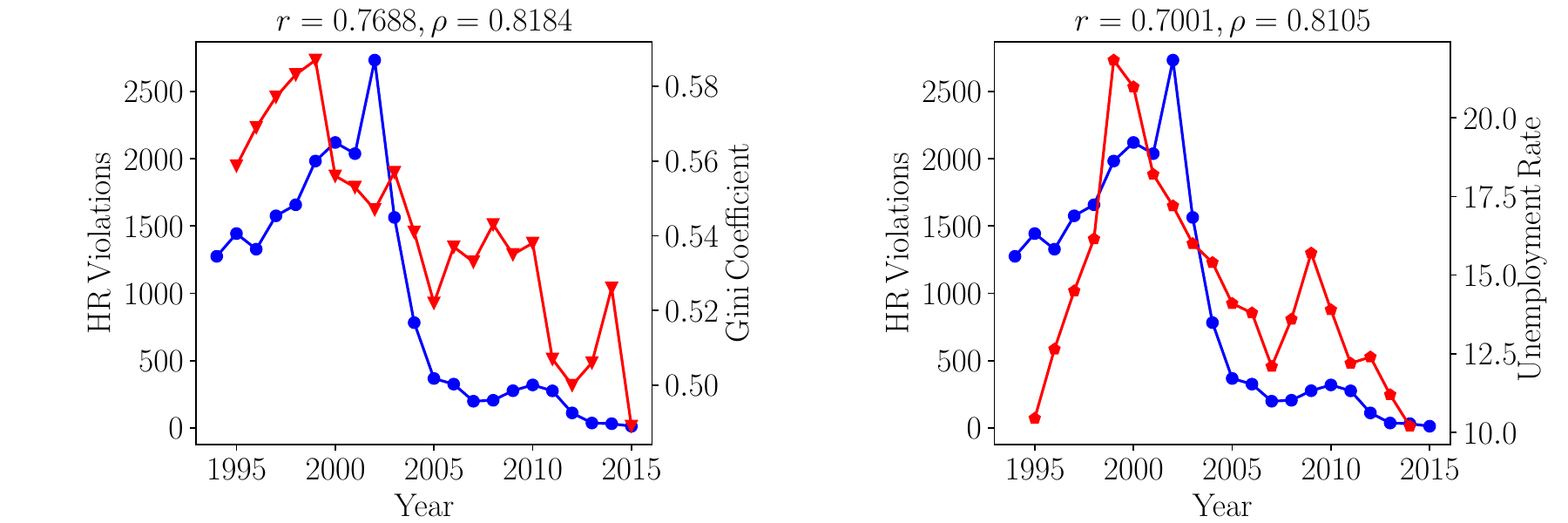}
   \end{tabular}
    \end{center}
   \caption{BLUE: Time series of the number of Human Rights Violations (HR Violations) in Medellin 
   during the years 1995-2015. 
   RED: (left panel) Gini Coefficient for Medellin and (right panel) Unemployment Rate  
   in Medellin between 1995-2015.}
    \label{fig:ViolentSoc}
\end{figure}

Fig~\ref{fig:ViolentSoc} depicts the time series of the numbers of violations of Human Rights in 
Medellin with two significant socio-economic factors: the Unemployment 
Rate (left panel) 
and the Gini Coefficient (right panel).
Linear, measured by the $r$-Pearson coefficient, as well as non-linear, measured by the $\rho$-Spearman 
coefficient, correlations are displayed in each panel.
Fig~\ref{fig:ViolentSoc} shows that the Unemployment Rate (UR) and the Gini Coefficient
are highly correlated, with the number of Human Rights Violations (HR Violations). 

For the data in Fig~\ref{fig:ViolentSoc}, the Unemployment Rate extracted from the reports of the 
\textit{Departamento Administrativo Nacional de Estadistica of Colombia} (DANE)'s. 
For 1995 to 2000, the annual Unemployment Rate was calculated as the arithmetic average of the 
quarterly reported unemployment rate. 
For 2001 to 2015, the annual annual Unemployment Rate was literally extracted from DANE's reports.
As to the Gini coefficient, DANE started reporting it only in 2002. In 2005, it changed its methodology 
and did no report data for 2006 and 2007.
The Gini coefficient time-series was completed as follows: data for 2001, 2006 and 2007 was calculated 
by using a linear regressor from the reported data (see, S1 Table for details) and the data from 1995 to 2000 
was obtained from reports of the World Bank for Colombia, which was additionally supplemented by a 
linear interpolation to get an approximate coefficient for 1996,1997 and 1998.

Since not all  Human Rights Violations can be exclusively associated to the gang conflict, 
the context analysis is complemented with the Homicide Rate per 100,000 inhabitants, the 
category that is mainly associated to the gang conflict.
These four variables comprises a fully-connected and strongly correlated network (see below), thus 
confirming the correlation between violence and socio-economic factors.
Focus here is on finding how the structure of the conflict network affects the behaviour of the gangs 
and violence indexes in Medellin. 
Specifically, how the number of gangs $N_\mathrm{g}$, the leading $\lambda_{\mathrm L}$ and second 
leading $\lambda_{\mathrm{L}_2}$ eigenvalues of the adjacency matrix correlate with the socio-economic
 and violence indexes.

\begin{figure}[!h]
  \begin{center}
    \begin{tabular}{c | c} \hline\hline
     \multicolumn{2}{c}{Pearson correlation network for $r\ge 0.6$ and $p\le 0.07$} \\ \hline
    \begin{minipage}[c]{0.425\textwidth} \includegraphics[width=\textwidth]{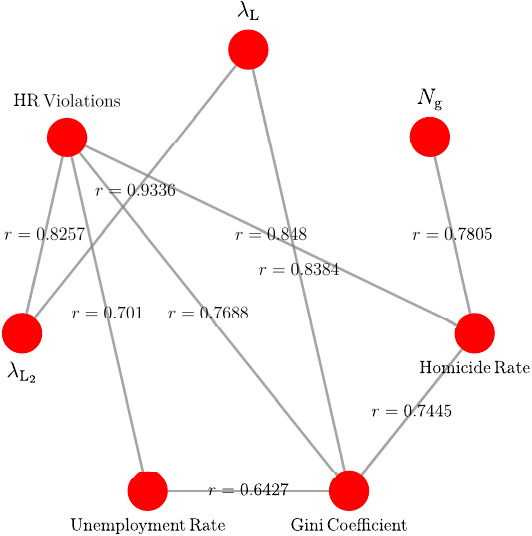} \end{minipage}
    &
    \begin{minipage}[c]{0.425\textwidth} \includegraphics[width=\textwidth]{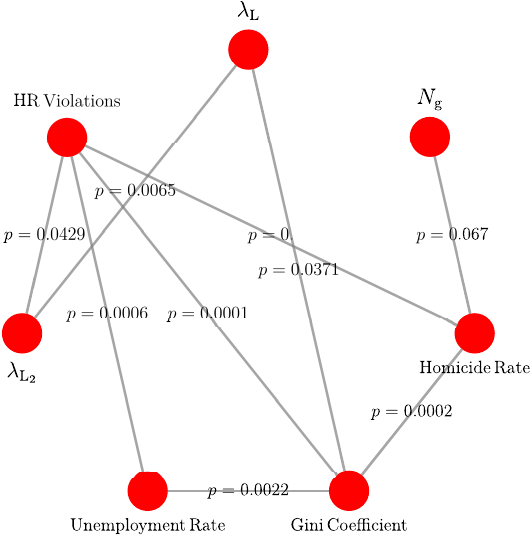} \end{minipage}
    \\ \hline\hline
    \multicolumn{2}{c}{Spearman correlation network for $\rho\ge 0.6$ and $p\le 0.07$} \\ \hline
    \begin{minipage}[c]{0.425\textwidth} \includegraphics[width=\textwidth]{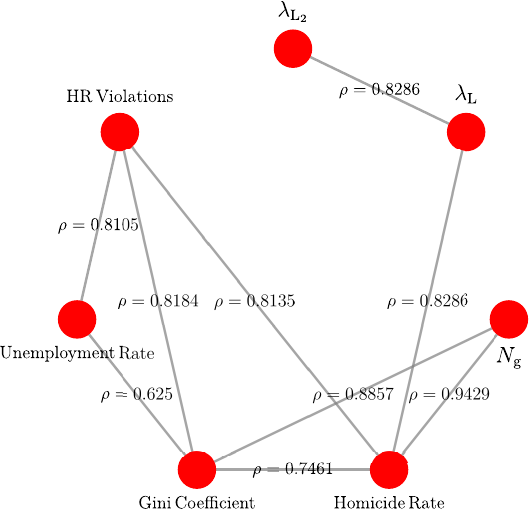} \end{minipage}
    &
    \begin{minipage}[c]{0.425\textwidth} \includegraphics[width=\textwidth]{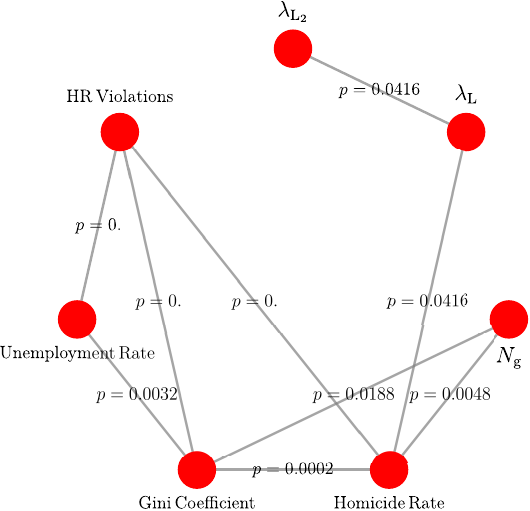} \end{minipage}
    \\ \hline\hline
   \end{tabular}
   \end{center}
     \caption{
 Linear and non-linear correlation networks of the number of Human Rights  Violations (HR Violations), 
 Homicide Rate (HR) per 100,000 inhabitants, Unemployment Rate, Gini  Coefficient, Number of gangs 
 $N_\mathrm{g}$ and the leading $\lambda_{\mathrm L}$ and second leading 
 $\lambda_{\mathrm{L}_2}$ eigenvalues of the adjacency matrix.
 For the same correlation network, left panels presents the $r,\rho$-values and right panels do for $p$-values.
     Only correlations satisfying $r,\rho\ge 0.6$ and $p\le 0.07$ were considered.}
    \label{fig:CorrFLLNTWRK}
\end{figure}
Fig~\ref{fig:CorrFLLNTWRK} presents the network of linear (top panel) and non-linear (bottom panel)
correlations of the confrontation network parameters with the number of Human Rights Violations (HR Violations), 
Homicide Rate (HR) per 100,000 inhabitants, Unemployment Rate and the Gini Coefficient.
Fig~\ref{fig:CorrFLLNTWRK} shows that non-linear correlations are stronger than linear ones. 
Specifically, note that the non-linear correlation network remains fully connected for $p\le 0.05$ 
whereas for that value of $p$, $N_\mathrm{g}$ is left as an isolated node in the linear correlation 
network.
However, the fact that each network presents correlations between different variables suggests that 
the correlation analysis should consider both type of correlations complementarily rather than independently.
Thus, Fig~\ref{fig:CorrFLLNTWRK} shows that the leading eigenvalue $\lambda_{\mathrm L}$ 
of the confrontation network linearly correlates with the Gini Coefficient and is non-linearly correlated 
with the Homicide Rate, the second leading $\lambda_{\mathrm{L}_2}$ linearly correlates with the number
of Human Right Violations whereas $N_\mathrm{g}$ has a weak linear correlation
but a strong non-linear correlation with the Homicide Rate and the Gini Coefficient. 

The correlation between the number of gangs and the Homicide Rate directly follows from the 
armed-domination strategy of paramilitary groups in the city: gangs in opposition and not subordinated 
to the hegemonic paramilitary structures were annihilated or decimated; thus, increasing the number
of homicides in the city.
In the framework of the Frobenius-Perron theorem, $\lambda_{\mathrm L}$ is interpreted as the
rate at which the number of confrontations increases or decreases and therefore of the number of 
homicides; this explains the correlation between $\lambda_{\mathrm L}$ and the Homicide Rate.
On an abstract ground, gangs that are highly connected to other highly connected gangs will incur 
in a higher probability of cascade violence.
The second leading eigenvalue $\lambda_{\mathrm{L}_2}$ can be interpreted as the rate of convergence 
to equilibrium distribution and it is associated to the mixing time of the network; this explains why 
it correlates with HR Violations, which not only incorporates assassination of gang members.
Fig~\ref{fig:CorrFLLNTWRK} thus allows to conclude that the structure of the confrontation network, 
the socio-economic variables and violence indexes in the city are certainly correlated. 

\subsection*{BLV Model for the Confrontation Network}
\begin{figure}[!h]
  \begin{center}
    \begin{tabular}{c c c}
    \small{1995-2000} & \small{2000-2002} & \small{2002} \\
    \includegraphics[width=0.3\textwidth]{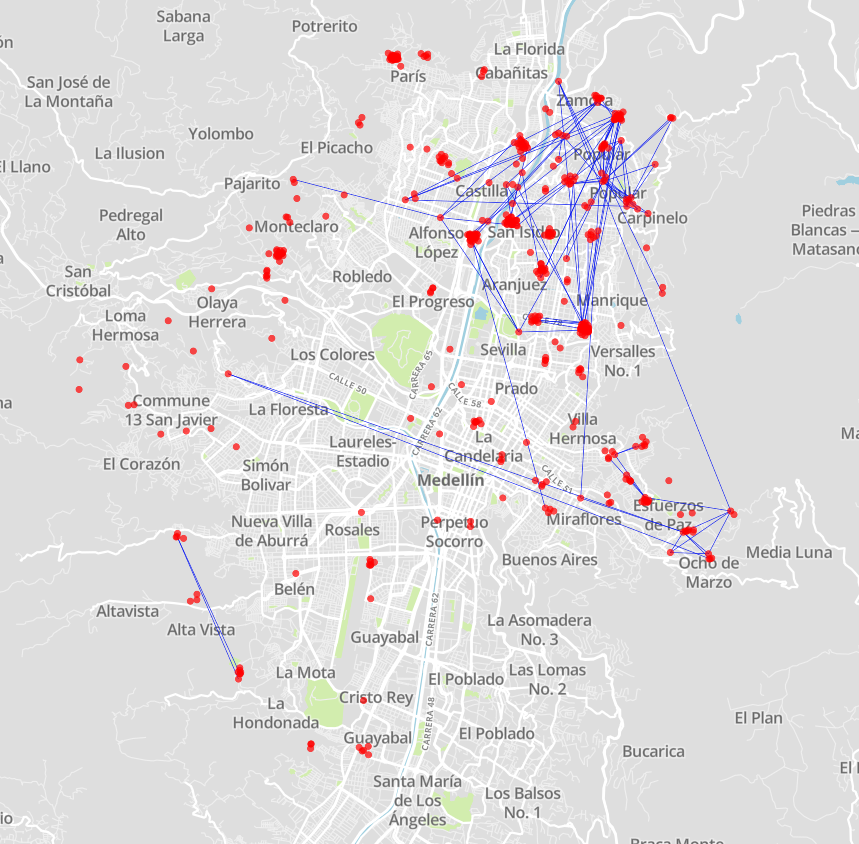} & 
    \includegraphics[width=0.3\textwidth]{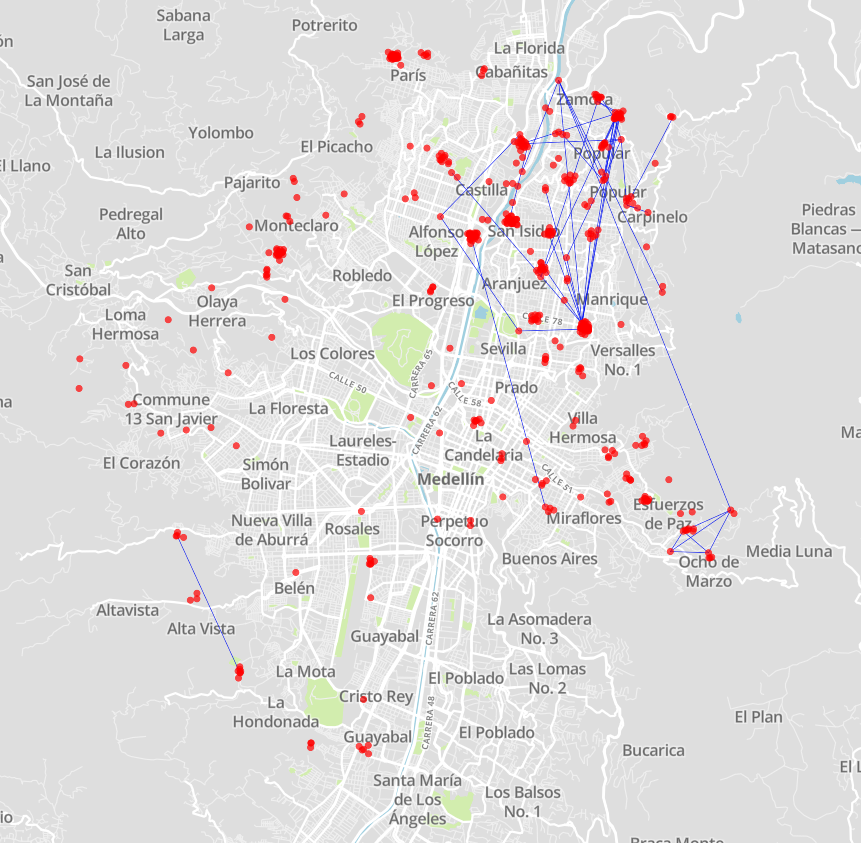} & 
    \includegraphics[width=0.3\textwidth]{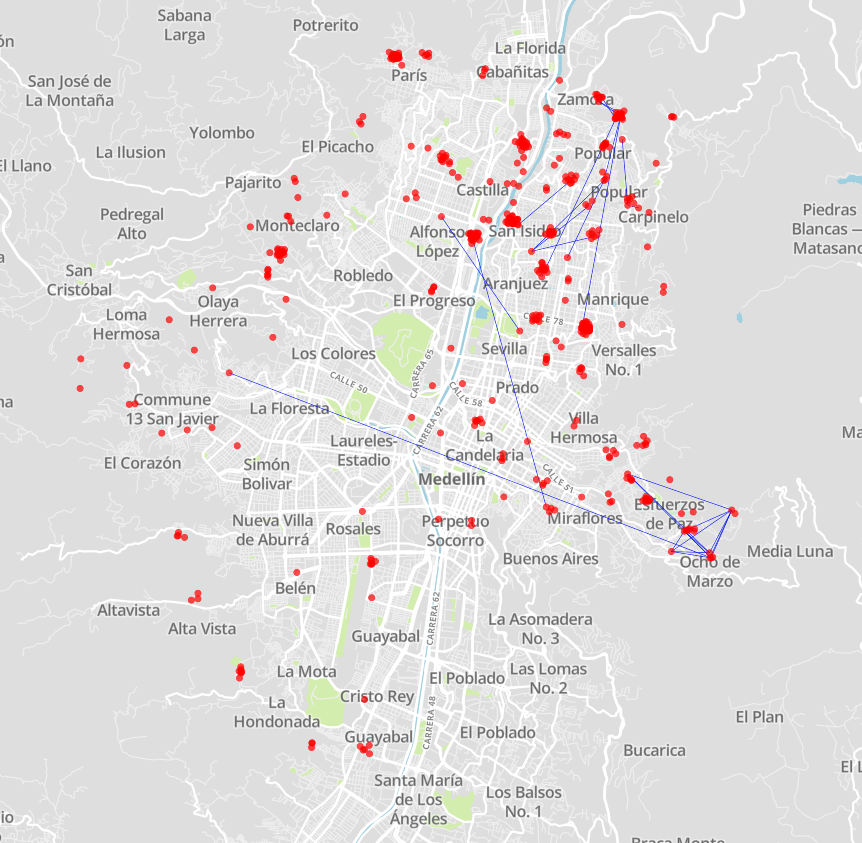}\\
      \small{2003} & \small{2003-2008} & \small{2008-2014} \\
     \includegraphics[width=0.3\textwidth]{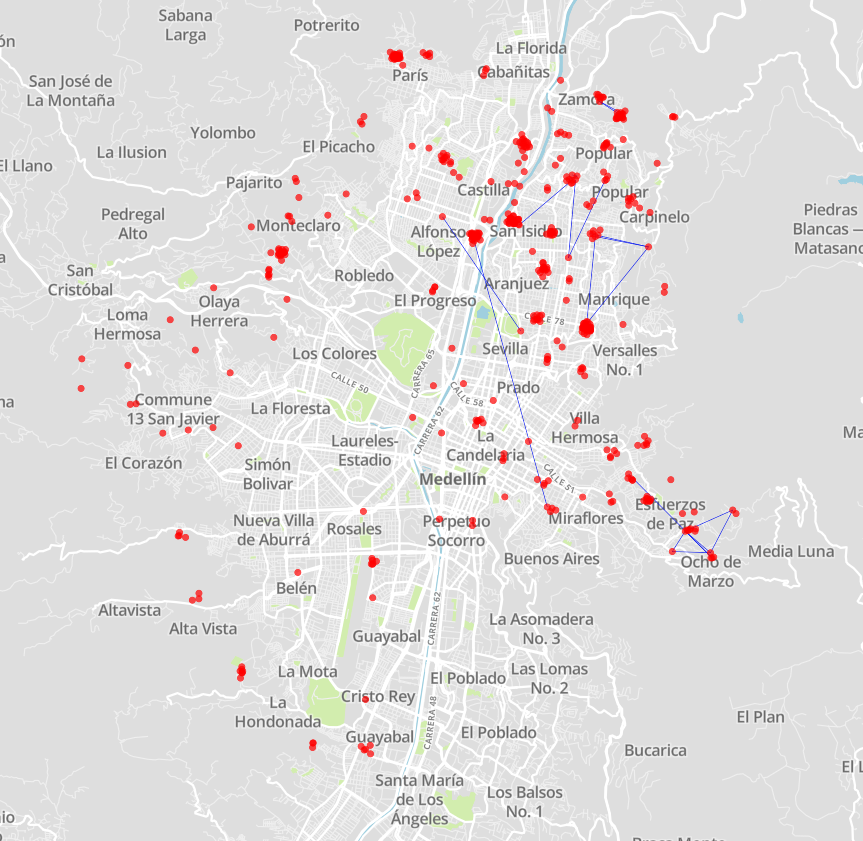}   &  
     \includegraphics[width=0.3\textwidth]{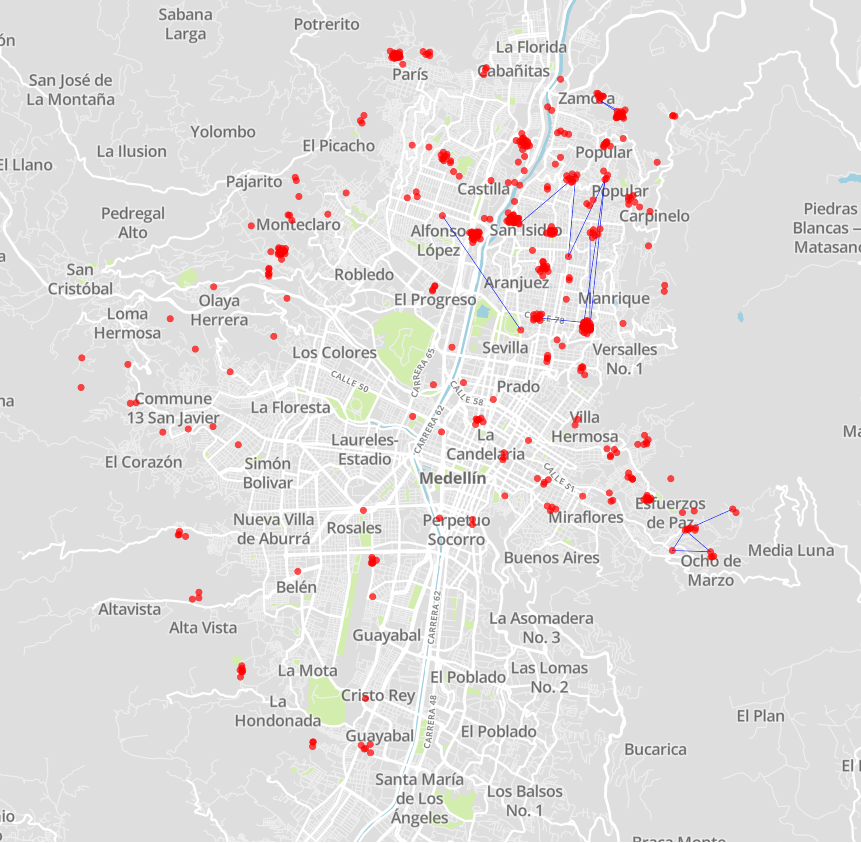} &  
     \includegraphics[width=0.3\textwidth]{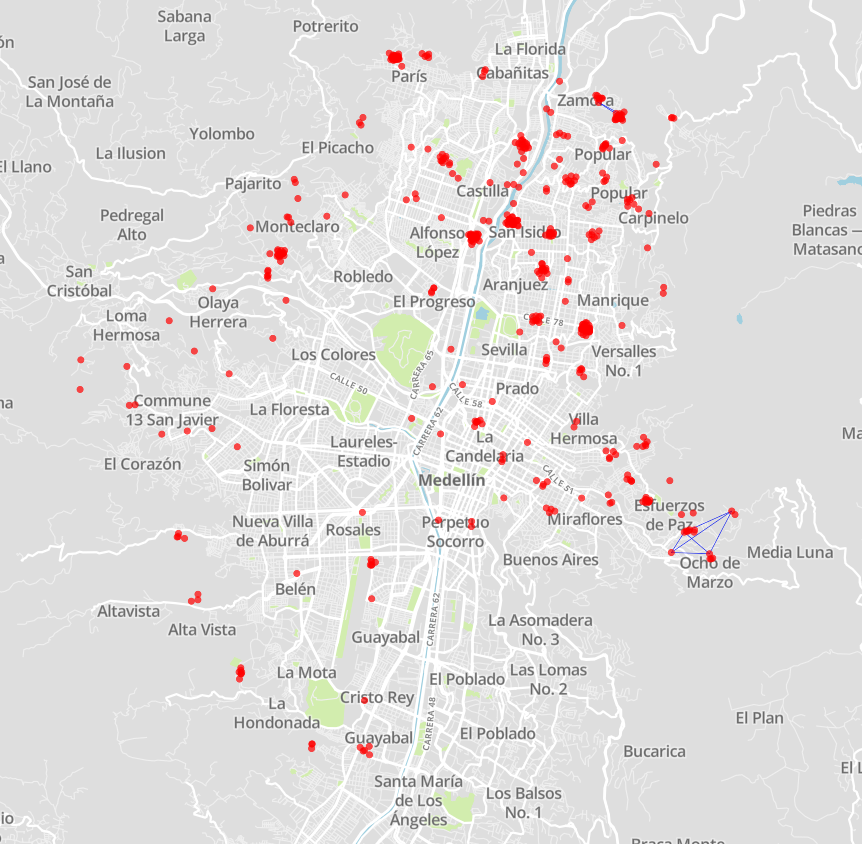}
   \end{tabular}
   \end{center}
 \caption{Gang confrontation network in Medellin. The localization of the gangs is presented as a red dot 
 in the map of Medellin and the enmity ties are represented by blue lines. 
 The background map was built using Mapbox.}
  \label{fig:conflict_networks}
\end{figure}

\begin{figure}[!h]
  \centering
  \includegraphics[width=0.8\textwidth]{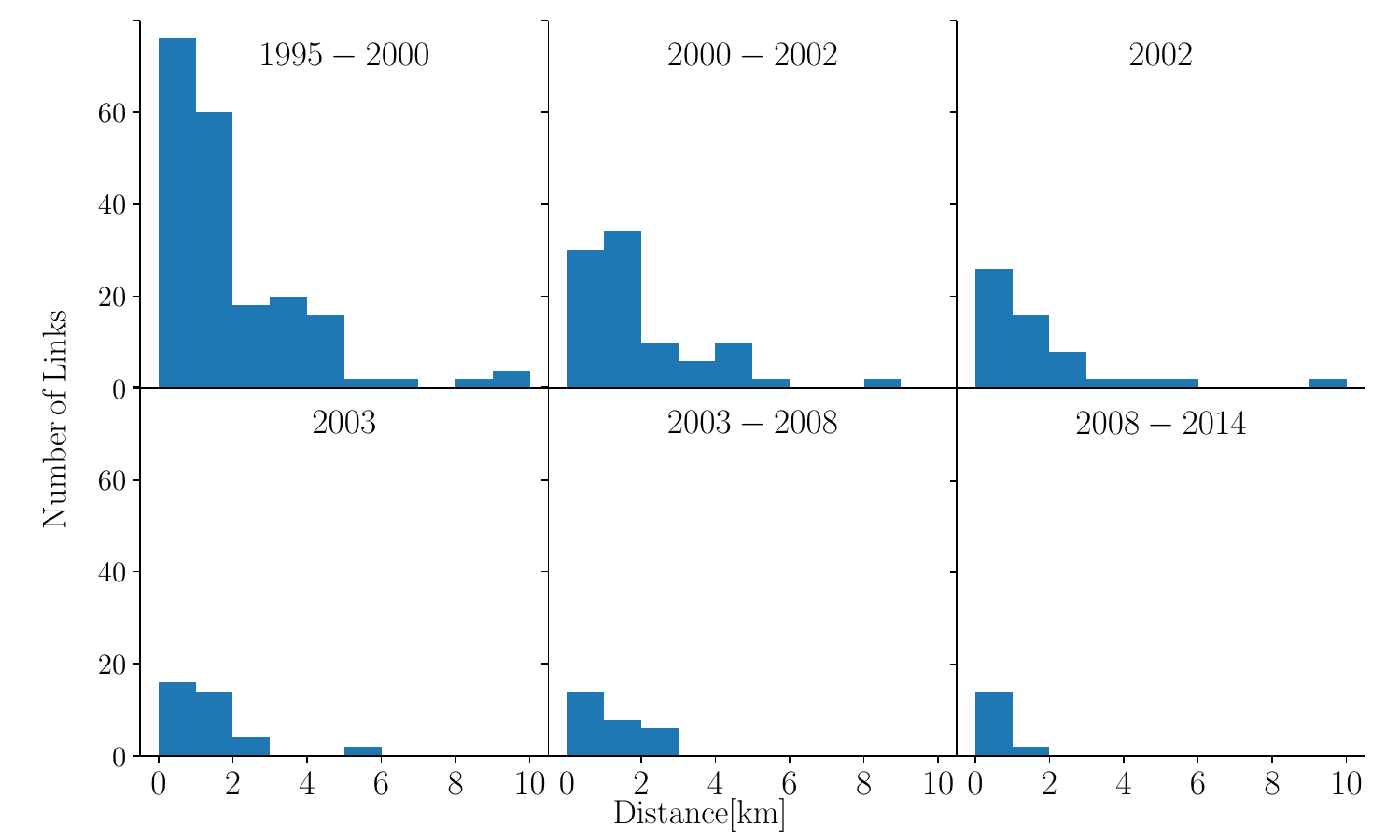}
\caption{Distribution of links based on the geographic distance between nodes.}
 \label{fig:DegHistAll}
\end{figure}

Fig~\ref{fig:conflict_networks} depicts the geographic location of gangs in the city by red dots 
and the hostility ties by blue lines between them. 
The network is highly space-correlated, i.e., the ties among the nodes are mainly between nearby 
nodes instead of distant nodes. 
This is also supported by the distribution of links based on the geographic distance between 
nodes presented in Fig~\ref{fig:DegHistAll}.
Therefore, the geolocalization of gangs is a significantly important factor to deeply 
understand their behaviour in the network. 
This is particularly significant since it points to the fact that the conflict relations of gangs in 
Medellin are in great proportion related to territory control, confirming the hypothesis previously 
proposed by other authors \cite{echeverria2000ciudad, melguizo2001evolution, sanin2004crime, 
alcala2006jovenes, restrepo2010medellin}.

\begin{figure}[!h]
  \begin{center}
    \begin{tabular}{c | c}\hline\hline
     \multicolumn{2}{c}{Pearson (l.h.s) and Spearman (r.h.s) correlation network for $r,\rho\ge 0.6$ and $p\le 0.07$} \\ \hline
    \begin{minipage}[c]{0.425\textwidth} \includegraphics[width=\textwidth]{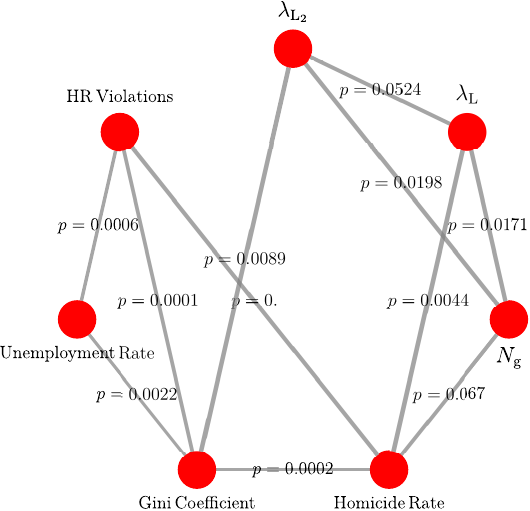} \end{minipage}
   &  
    \begin{minipage}[c]{0.425\textwidth} \includegraphics[width=\textwidth]{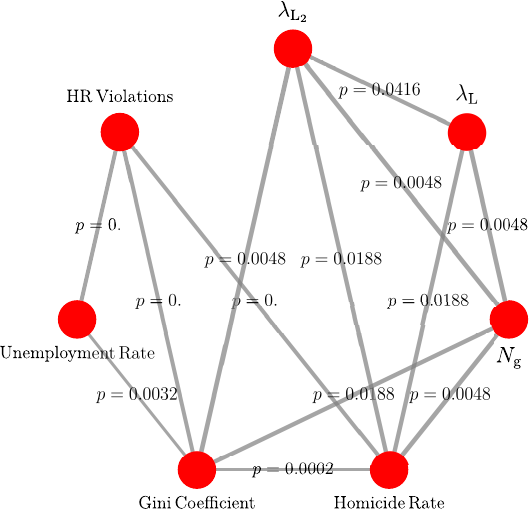} \end{minipage}
        \\ \hline\hline
   \end{tabular}
   \end{center}
     \caption{
 Linear and non-linear correlation networks of the number of Human Rights Violations (HR Violations), 
 Homicide Rate (HR) per 100,000 inhabitants, Unemployment Rate, Gini  Coefficient, Number of gangs 
 $N_\mathrm{g}$ and the leading $\lambda_{\mathrm L}$ and second leading 
 $\lambda_{\mathrm{L}_2}$ eigenvalues of the adjacency matrix.
     Edges' thickness is proportional to $r,\rho$ coefficients (the thickest
     edge corresponds to the $r=0.943$ correlation between the Gini Coefficient and $N_g$ whereas the 
     thinest edge corresponds to $r=0.625$ correlation between the Gini Coefficient and the Unemployment 
     Rate.
    Only correlations satisfying $r,\rho\ge 0.6$ and $p\le 0.07$ were considered.}
    \label{fig:CorrGEORFRNCD}
\end{figure}

Therefore, the geolocalisation of gangs is a significantly important factor to deeply 
understand their behaviour in the network and motivates the usage of spatial network 
models \cite{barthelemy2011spatial}.
Specifically, {the fact that gangs operate 
under cost-benefit principles} suggests that the BLV formalism can
be a sensible approach to model Medellin's gang phenomenon; {however, the 
non-stationary character of Medellin's violence may undermine its applicability.}
Since the number of nodes {for which confrontation information was available (blue lines
in Fig~\ref{fig:conflict_networks})} decreases in the geo-referenced network and the graph is considered 
unweighted (see above), then it is instrumental to recalculate the correlation network, results are presented in 
Fig~\ref{fig:CorrGEORFRNCD} for linear (l.h.s panel) and non-linear (r.h.s panel) correlation measures.

In contrast to Fig~\ref{fig:CorrFLLNTWRK}, Fig~\ref{fig:CorrGEORFRNCD} shows that the
context variables (HR Violations, Homicide Rate, Unemployment Rate and Gini Coefficient) 
and confrontation network properties ($N_\mathrm{g}$, $\lambda_{\mathrm L}$ and 
$\lambda_{\mathrm{L}_2}$) aggregate in well defined fully-connected-closed-networks.
For linear correlations, ``interaction" between closed networks is provided by the correlation 
between $\lambda_{\mathrm L}$ and $\lambda_{\mathrm{L}_2}$ with the Homicide Rate and
Gini Coefficient, respectively.
For non-linear correlations, ``interaction" between closed networks is provided by the additional 
correlation between $\lambda_{\mathrm{L}_2}$ and Homicide Rate.
Despite the differences between the full and the geo-referenced confrontation-networks, and 
therefore of the particular correlations between variables, the strong non-linear correlation between 
$\lambda_{\mathrm L}$ and the Homicide Rate is common to both confrontation-networks.
This is a robust and key finding of the present work.

\subsubsection*{BLV-Predicted Adjacency Confrontation Matrix}
\begin{figure}[!h]
  \begin{center}
   \begin{tabular}{c c c}
    \small{1995-2000} & \small{2000-2002} & \small{2002}\\
    \includegraphics[width=0.29\textwidth]{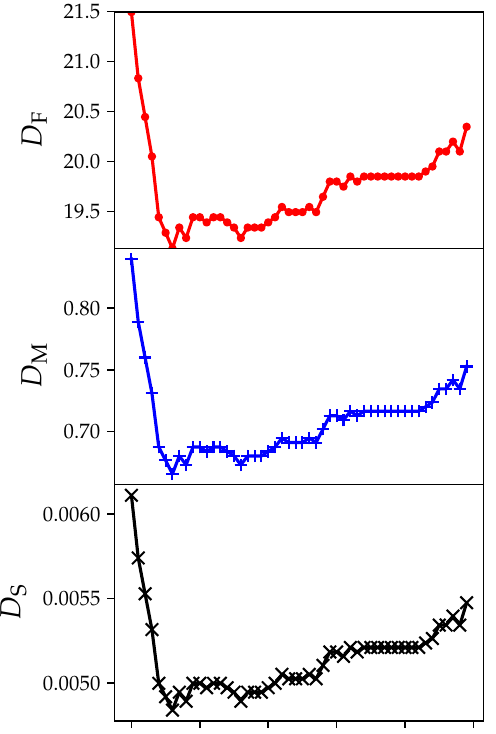} &  
    \includegraphics[width=0.29\textwidth] {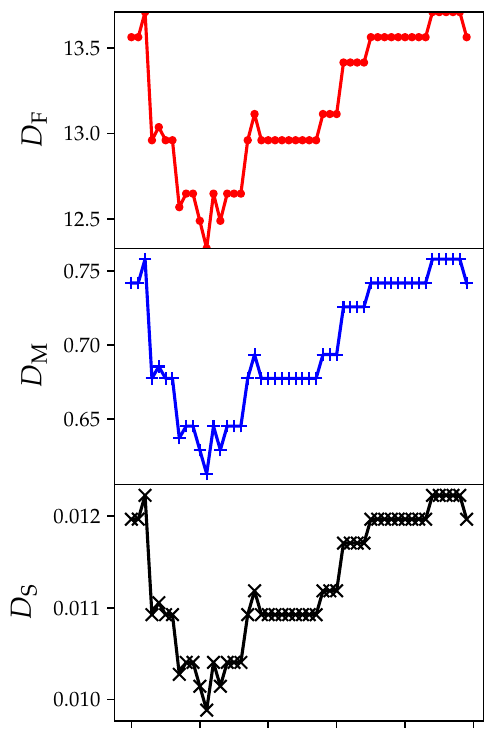} & 
    \includegraphics[width=0.29\textwidth]{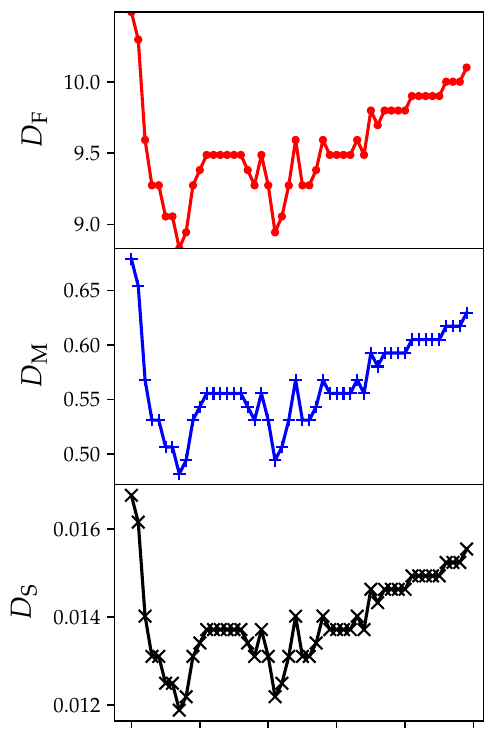}
    \\    
    \small{2003} & \small{2003-2008} & \small{2008-2014} \\
     \includegraphics[width=0.29\textwidth]{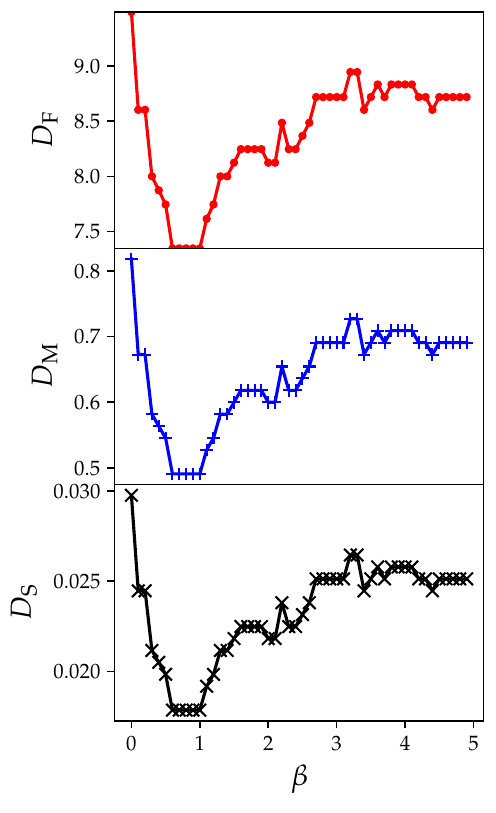} & 
     \includegraphics[width=0.29\textwidth]{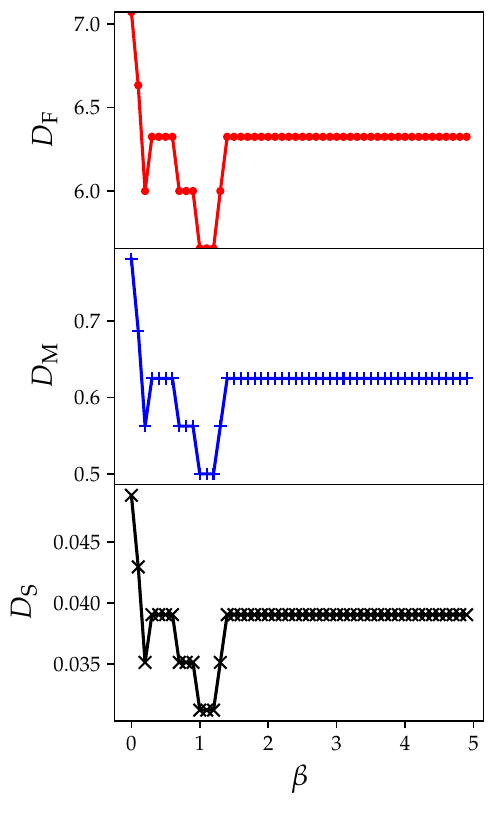} &  
     \includegraphics[width=0.29\textwidth]{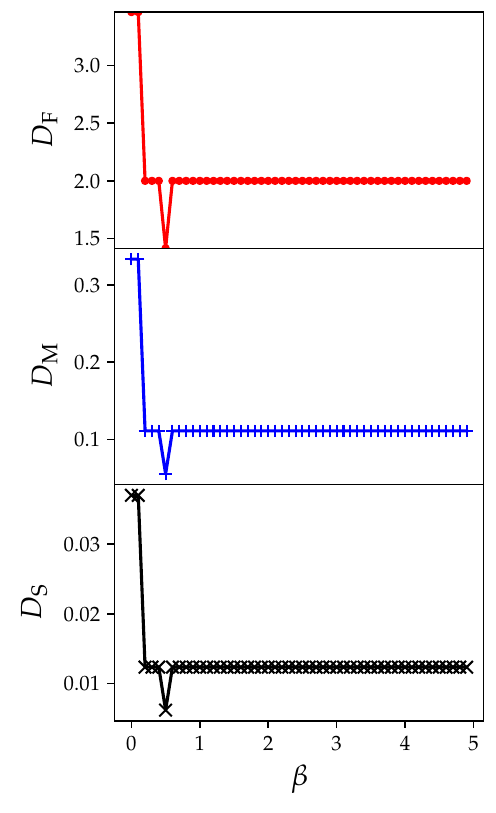}\\
   \end{tabular}
   \end{center}
 \caption{Matrix distance $D_\mathrm{F}$, $D_\mathrm{M}$ and $D_\mathrm{S}$ 
  as a function of $\beta$ parameter.
  For each period of time, optimal the value corresponds to $\beta = \{0.6, 1.1, 0.7, 0.6, 1.0, 0.5\}$.
  A high value of $\beta$ represents a short distance interaction and conversely, a low value implies 
  long-range interaction.
 }
  \label{fig:optimization}
\end{figure}

For each period of time, the values of the three distance measures defined above
are shown in Fig~\ref{fig:optimization}.
The Multiplying distance $D_\mathrm{M}$ is bounded: $0\le D_\mathrm{M} \le 1$ 
ranging between identical matrices  to matrices having orthogonal support, 
respectively.
Since adjacency matrices considered here are not normalized, then it is not straightforward 
to assign a  particular interpretation to a given value of the distance measures. 
However, note that the behavior of the three measures, as a function of $\beta$, is identical.
Therefore, despite that high spatial character of the confrontation network in 
Figs~\ref{fig:DegHistAll} and \ref{fig:conflict_networks}, Fig~\ref{fig:optimization} shows that 
the BLV method does not accurately described the reconstructed adjacency matrix 
since $D_\mathrm{M} \sim 0.5$.
This can be a consequence of the small dataset in the present version of the reconstructed 
network or, {as anticipated above,} an indication that the BLV may not applied 
 {to Medellin's scenario due to the lack of stationary 
character of confrontations.}
Different alternatives to quantitatively predict the spatial embedding of the conflict are currently 
under study.
%

\section*{Conclusions}
The analysis presented in Fig~\ref{fig:resilience_networks} allows concluding that the conflict 
networks are resilient against removing random nodes, but weak against the targeted 
deletion of high–degree nodes. 
Figs~\ref{fig:CorrFLLNTWRK} and \ref{fig:CorrGEORFRNCD} clearly show that the 
socio-economic variables utilized here and the confrontation network properties are moderate 
to highly correlated to the escalation of conflict in Medellin during the last twenty years.
Particularly, the correlation among the leading eigenvalue of the adjacency matrix 
$\lambda_\mathrm{L}$ and the Homicide Rate allows to conclude that 
the topological structure of the network is a significant descriptor of the violence in the city. 
Since the leading eigenvalue is always higher than two, based on the Frobenuios-Perron 
theorem, it can be concluded that the conflict network does not reach stability. 
This can be interpreted as a measurement of the instability of the conflict network, 
leading to retaliation among gangs and hence manifested as an occurrence on the number 
of Human Rights Violations in the city and vice-versa.

Figs~\ref{fig:DegHistAll} and \ref{fig:conflict_networks} suggest that the conflict network 
of gangs in Medellin is spatial.
However, the BLV formalism do not accurately describe the reconstructed confrontation network
and different alternatives to quantitatively predict the spatial embedding of the conflict are 
currently under study.
To summarize, a high correlation between the structure of a gang confrontation network 
and the escalation of conflict in Medellin was presented and that the territory control mechanism 
is a main driven force of the conflict among the gangs in Medellin.

\section*{Acknowledgements}
JDB and GAA-G thank COLCIENCIAS for their support through a doctoral scholarship from Programs No. 727 
and Pasaporte a la Ciencia, respectively.
WG acknowledges funding from The Alan Turing Institute through an EPSRC grant EP/N510129/1 and DSTL 
project ACC6005162.
LAP acknowledges support from the Dean of \textit{Facultad de Ciencias Exactas y Naturales}  and 
by the \textit{Comit\'e para el Desarrollo de la Investigaci\'on} --CODI-- of Universidad de Antioquia, Colombia 
and under the \textit{Estrategia de Sostenibilidad}.

\section*{Supporting information}

\indent \textbf{S01 Table. Socio-economic data}. Dataset of socio-economic variables and matrix properties.

\textbf{S01 Script. Mathematica 11.0 Script.}
This Mathematica 11.0 script generates the results presented in Figs~\ref{fig:CPFAdjMat} and \ref{fig:resilience_networks} 
and Tables~\ref{tab:fitCPF} and \ref{tab:prmtrsfitCPF} of the manuscript.

\textbf{S02 Script. Mathematica 11.0 Script.}
This Mathematica 11.0 script generates the results in Figs~\ref{fig:CorrFLLNTWRK} and \ref{fig:CorrGEORFRNCD}
of the manuscript.

\textbf{S01 DataFile. Mathematica 11.0 Data File.} Contains the node name of each gang.

\textbf{S02 DataFile. Mathematica 11.0 Data File.} Contains the edges of the confrontation network before the Bloque Metro period.

\textbf{S03 DataFile. Mathematica 11.0 Data File.} Contains the edges of the confrontation network during the Bloque Metro period.

\textbf{S04 DataFile. Mathematica 11.0 Data File.} Contains the edges of the confrontation network during the Bloque Metro and Bloque Cacique Nutibara Confrontation period.

\textbf{S05 DataFile. Mathematica 11.0 Data File.} Contains the edges of the confrontation network during the Bloque Cacique Nutibara period.

\textbf{S06 DataFile. Mathematica 11.0 Data File.} Contains the edges of the confrontation network during the Bloque Heroes de Granada period.

\textbf{S07 DataFile. Mathematica 11.0 Data File.} Contains the edges of the confrontation network during the DDR period.



\end{document}